\documentclass[10pt,a4paper]{article}
\RequirePackage{latexsym,amsmath,amssymb}
\RequirePackage[dvipsnames,usenames]{color}
\usepackage{cite}
\usepackage{fullpage}

\usepackage[british]{babel}
\usepackage[latin1]{inputenc}
\usepackage[T1]{fontenc}
\usepackage[final]{showkeys} 
\usepackage{hyperref}
\hypersetup{pdfauthor={M. Cadoni, M. Tuveri...},
            pdftitle={}
            }
\usepackage[]{cleveref}
\Crefname{equation}{Eq.}{Eqs.}
\Crefname{figure}{Fig.}{Figs.}
\crefformat{plural}{#2eqs.~(#1)#3}
\crefname{section}{Sect.}{Sects.}

\usepackage{amsfonts}
\newcommand{\be}{\begin{eqnarray}}
\newcommand{\ee}{\end{eqnarray}}

\renewcommand{\d}{\mbox{${\rm d}$}} 
\newcommand{\lp}{\ell_{\rm p}}
\newcommand{\mpl}{m_{\rm p}}
\newcommand{\Ng}{N_{\rm G}}
\newcommand{\eg}{\varepsilon_{\rm G}}
\newcommand{\Nb}{N_{\rm B}}
\newcommand{\NL}{N_{\Lambda}}

\newcommand{\gn}{G_{\rm N}}

\newcommand{\Rh}{R_{\rm H}}
\newcommand{\Rb}{R_{\rm B}}
\newcommand{\mb}{m_{\rm B}}

\newcommand{\DE}{{\rm DE}}
\newcommand{\DF}{{\rm DF}}
\newcommand{\ab}{a_{\rm B}}

\newcommand{\ppar}{{p_\parallel}}

\newcommand{\Neff}{N_{\text{eff}}}
\newcommand{\rmd}{\mathrm{d}}
\newcommand{\e}[1]{\operatorname{e}^{#1}}
\makeatletter\g@addto@macro\bfseries{\boldmath}\makeatother%
\DeclareSymbolFont{lettersA}{U}{txmia}{m}{it}
	\DeclareMathSymbol{\R}{\mathord}{lettersA}{"92}
	\DeclareMathSymbol{\C}{\mathord}{lettersA}{"83}
\title{\bf Emergence of a Dark Force in Corpuscular Gravity}
\author{M.~Cadoni${}^{ab}$\thanks{E-mail: mariano.cadoni@ca.infn.it},
R.~Casadio${}^{cd}$\thanks{E-mail: casadio@bo.infn.it},
A.~Giusti${}^{cde}$\thanks{E-mail: agiusti@bo.infn.it},
and 
M.~Tuveri${}^{ab}$\thanks{E-mail: matteo.tuveri@ca.infn.it}
\\
\\
${}^a$\emph{Dipartimento di Fisica, Universit\`a di Cagliari}
\\
{\em Cittadella Universitaria, 09042 Monserrato, Italy}
\\
\\
${}^b$\emph{I.N.F.N, Sezione di Cagliari, Cittadella Universitaria, 09042 Monserrato, Italy}
\\
\\
${}^c$\emph{Dipartimento di Fisica e Astronomia, Universit\`a di Bologna}
\\ 
{\em via Irnerio~46, 40126 Bologna, Italy}
\\
\\
${}^d$\emph{I.N.F.N., Sezione di Bologna, IS - FLAG}
\\
{\em via B.~Pichat~6/2, 40127 Bologna, Italy}
\\
\\
${}^e$\emph{Arnold Sommerfeld Center, Ludwig-Maximilians-Universit\"at}
\\
{\em Theresienstra{\ss}e~37, 80333 M\"unchen, Germany}
\\
\\
}
\begin{document}
\maketitle
\begin{abstract}
We investigate the emergent laws of gravity when Dark Energy and 
the de~Sitter space-time are modelled as a critical Bose-Einstein 
condensate of a large number of soft gravitons $N_{\rm G}$.
We argue that this scenario requires the presence of various regimes 
of gravity in which $N_{\rm G}$ scales in different ways.
Moreover, the local gravitational interaction affecting baryonic 
matter can be naturally described in terms of gravitons pulled out 
from this Dark Energy condensate (DEC). We then explain the additional 
component of the acceleration at galactic scales, commonly attributed 
to dark matter, as the reaction of the DEC to the presence of baryonic 
matter. This additional dark force is also associated to gravitons 
pulled out from the DEC and correctly reproduces the MOND acceleration.
It also allows for an effective description in terms of General 
Relativity sourced by an anisotropic fluid.
We finally calculate the mass ratio between the contribution of the 
apparent dark matter and the baryonic matter in a region of size $r$ 
at galactic scales and show that it is consistent with the 
$\Lambda$CDM predictions.
\end{abstract}
%
%
%
%
%
%
%
%
%
\section{Introduction}
\label{Intro}
One of the major ideas triggering recent theoretical progress about the 
gravitational interaction is that of {\sl emergent gravity}: the classical 
space-time structure and gravity emerge together from an underlying  
microscopic quantum theory~\cite{Sakharov:1967pk,Padmanabhan:2014jta,Verlinde:2016toy}.
The power of this emergent paradigm is that it must depend loosely on the 
details of the underlying microscopic theory and it is essentially determined 
by its fundamental quantum nature. 
\par
The notion of emergent gravity is quite general and it has been used in several 
different contexts~\cite{Sakharov:1967pk,Padmanabhan:2009vy,Padmanabhan:2014jta,Padmanabhan:2016eld,
Dvali:2011aa,Dvali:2013eja,Jacobson:1995ab,Bhattacharyya:2008jc,Volovik:2000ua}. 
In this paper we want to focus on two main realisations of this idea, which 
have recently attracted a lot of attention. The first one uses the {\sl entanglement\/} 
of microscopic quantum states as the origin of space-time geometry.
This route historically starts from the discovery of the Bekenstein-Hawking~(BH) 
entropy area law~\cite{Bekenstein:1973ur}, goes trough the development of the 
AdS/CFT correspondence~\cite{Maldacena:1997re} and the Ryu-Takayanagi formula, 
where it clearly appears that the BH formula is related to the quantum entanglement 
of the vacuum~\cite{Ryu:2006bv}. Subsequently, it was also realized that  
quantum entanglement could explain the connectivity of classical 
space-time~\cite{VanRaamsdonk:2010pw}, and that the linearized Einstein 
equations can be derived from quantum information principles~\cite{Jacobson:2015hqa}.
The second main realization of the idea that gravity is emergent uses the 
notions of {\sl quantum compositeness\/} and {\sl classicalization\/}~\cite{Dvali:2011aa,Dvali:2013eja}. 
Gravitational systems, such as black holes and cosmological spaces, can be 
described as a composite quantum system of a large number $N_{\rm G}$ of soft 
gravitons.
It has been shown that these gravitational systems exhibit properties of a 
Bose-Einstein condensate (BEC) at the quantum critical point. Moreover, 
the usual classical space-time structure emerges in the limit 
$N_{\rm G}\to\infty$ of this picture~\cite{Dvali:2011aa,Dvali:2013eja}. 
This corpuscular realization of the paradigm of emergent gravity has 
been also successfully used to describe Hawking radiation~\cite{Dvali:2012rt,Casadio:2015lis} 
and inflation~\cite{Dvali:2013eja,Casadio:2017twg,Berezhiani:2016grw}.
\par
One common intrinsic feature of both the above realizations of emergent 
gravity is that of {\sl holography}. The intrinsic holographic nature of 
the quantum entanglement approach is evident in the Ryu-Takayanagi derivation 
of the entanglement entropy and, more in general, in the quantum
information picture of black holes and cosmological horizons~\cite{Ryu:2006bv,Verlinde:2016toy}.
The same holographic nature is at the heart of the corpuscular approach, which 
is based on the fact that the number $N_{\rm G}$ of soft gravitons in a BEC 
at the critical point scales, in terms of the size $r$ of the system, 
as~\footnote{We shall use units with $c=1$ but display explicitly the 
Planck constant $\hbar=\lp\,\mpl$ and Newton constant $\gn=\lp/\mpl$.},
\be\label{hb}
N_{\rm G}
\sim
\frac{r^2}{\lp^2}
\ ,
\ee
where $\lp$ is the Planck length.
The holographic character of this description is also very important 
for understanding the quantum information counted by the BH entropy
(see the complementary, firewall and ER=EPR discussion~\cite{Almheiri:2012rt,Maldacena:2013xja}) and 
for the description of black holes as BEC of gravitons at the 
critical point~\cite{Dvali:2011aa,Dvali:2013eja}. 
\par
The emergent gravity scenario also provides a connection between the 
microscopic ultraviolet (UV) scale $\lp$ and the infrared (IR) 
cosmological scale $L=H^{-1}$ of gravity (here $H$ is the Hubble 
parameter and $L$ the Hubble scale).
In the quantum entanglement setup, the entropy associated to the de~Sitter (dS)
space-time can be explained, similarly to the BH entropy, as a long 
range entanglement connecting bulk excitations with the dS horizon~\cite{Verlinde:2016toy}.
In the corpuscular setup, both black holes and our observable 
universe are ``maximally classical'' systems, {\em i.e.}~BEC at the 
critical point satisfying the relation~\eqref{hb} with
$r=\Rh$ (the Schwarzschild radius) and $r=L$, respectively. 
\par  
The success of emergent gravity in describing the holographic regimes 
of gravity shown in Eq.~\eqref{hb}, {\em i.e.}~black holes and the dS universe, 
gives a strong motivation to use it also at intermediate scales, {\em i.e.}~at
galactic scales.
It is quite evident that the behaviour of the gravitational interaction at these scales
cannot be simply described by the maximally packing condition~\eqref{hb}.
On the other hand, explaining the phenomenology of gravity at galactic scales
has been one of the main motivations for introducing dark 
matter~\cite{Zwicky:1933gu,Faber:1979pp,deSwart:2017heh}
and the $\Lambda$CDM model~\cite{Peebles:2002gy,Riess:1998cb,Penzias:1965wn,Ade:2013zuv}.
One is therefore led to expect that the application of the emergent gravity
scenario at galactic scales may hold the key for understanding the dark matter
mystery.
\par
In a fully emergent gravity scenario, such as the one we consider in this 
work, in which matter and space-time are intimately related, the existence 
of a form of matter different from the baryonic one is conceptually weird.
Moreover, we recall the $\Lambda$CDM model is not completely satisfactory 
also from the observational point of view, both at the level of galaxies and
galaxy clusters~\cite{Klypin:1999uc,Moore:1999nt,BoylanKolchin:2011de,BoylanKolchin:2011dk}.  
If one does not assume the existence of dark matter, a crucial challenge 
for every model of emergent gravity is the explanation of galaxy rotation 
curves~\cite{Rubin:1980zd,Persic:1995ru} and the Tully-Fisher relation~\cite{Tully:1977fu} 
between the velocity of stars far away from the galactic center and the
total baryonic mass $\mb$ contained in the galaxy.  
\par
In the framework of Modified Newtonian dynamics~(MOND)~\cite{Milgrom:1983ca,Milgrom:2014usa}, 
the Tully-Fisher relation is explained assuming that, at distances outside 
the galaxy's inner core, the gravitational acceleration experienced by a 
test particle is given by
\begin{equation}
\label{MOND}
a_{\rm MOND}(r)
=
\sqrt{\, \frac{a_{\rm B}(r)}{6\, L}}
\ ,
\end{equation}
where
\be
a_{\rm B}(r)
=
\frac{\gn\, \mb(r)}{r^2}
\label{aN}
\ee
is (minus) the Newtonian radial acceleration that would be caused by the 
baryonic mass $\mb=\mb(r)$ inside the radius $r$.
\par
A first step to explain Eq.~\eqref{MOND} in the framework of gravity emerging
from quantum entanglement has been undertaken in Ref.~\cite{Verlinde:2016toy}. 
In that work, it was shown that, when applied at galactic scales, the laws of 
emergent gravity contain an additional dark gravitational force, which 
may explain the phenomenology commonly attributed to dark matter and 
reproduce the MOND acceleration~\eqref{MOND}.
Following Verlinde~\cite{Verlinde:2016toy}, the long range entanglement 
connecting bulk excitations with the dS horizon ({\em i.e.}~the positive dark 
energy) generates a (thermal) volume contribution to the entanglement entropy 
and a subsequent competition between area and volume laws.
This can be seen as an elastic response of the dark energy medium to the 
presence of baryonic matter which, in turn, implies an additional dark 
gravitational force correctly reproducing the MOND acceleration~\cite{Verlinde:2016toy}.
\par
The purpose of this paper is to explain the generation of an additional 
dark gravitational force at galactic scales and derive the MOND acceleration~\eqref{MOND}
using the corpuscular approach to emergent gravity. 
This will be done by developing further and generalizing some ideas 
presented in Ref.~\cite{Cadoni:2017evg}, where an effective fluid approach 
for the Dark Energy condensate (DEC) of soft gravitons permeating the universe
has been used. 
\par
We shall begin with a critical discussion of various regimes of gravity in the 
corpuscular scenario.
In Section~\ref{sec2}, we will start by arguing that, describing dark energy 
as a critical BEC of soft gravitons (the DEC) implies not only the presence 
of a non-extensive regime of gravity satisfying Eq.~\eqref{hb}, but also of 
an extensive regime in which $N_{\rm G}\sim r^3/(L\,\lp^2)$.  The local 
gravitational interaction with baryonic matter can then be naturally described
in terms of gravitons pulled out from the DEC. We will first consider, 
in Section~\ref{sec3}, baryonic matter in the diluted approximation, when 
the local reaction of the condensate to the presence of baryonic matter can 
be (ideally) neglected. We will then proceed by describing what happens when 
we go beyond the diluted approximation and baryonic matter begins to clump.
We will show that, in this regime, the reaction of the DEC to the presence 
of baryonic matter is also associated to gravitons pulled out from the DEC. 
They generate an additional gravitational dark force on baryonic probe sources,
which correctly reproduces the MOND acceleration and allows for an effective
description in terms of General Relativity (GR) sourced by an anisotropic 
fluid. In Section~\ref{sec4}, we will also compute the ratio between the 
apparent dark matter and the baryonic component to the mass density and 
show that it is consistent with the $\Lambda$CDM result for the present 
abundance of different kinds of matter. We finally conclude with some 
considerations about future developments in Section~\ref{sec5}.  
\section{Quantum Compositeness and the scaling of graviton number}  
\label{sec2}
\setcounter{equation}{0}
Our starting point is that the true quantum nature of gravity cannot 
be fully neglected in our present universe, even at astrophysical and 
cosmological scales, and the geometric description given by
Einstein gravity (or modifications thereof) should only emerge in 
suitable regimes and for specific observables.
In particular, it has been conjectured that the quantum state of our 
universe could be thought of as a BEC~\cite{Dvali:2013eja} 
containing a certain number $\Ng$ of (very soft and virtual) gravitons 
with typical energy $\eg$, very much like the gravitational field 
of a black hole~\cite{Dvali:2011aa}. The presence of baryonic matter must 
affect the quantum state of this BEC of gravitons and, at least in some crude
approximation, one can then expect an energy balance, akin to the 
Hamiltonian constraint of GR, holds in the form
\be\label{hc}
H_{\rm B}+H_{\rm G}
=
0
\ ,
\label{H0}
\ee
where $H_{\rm B}$ is the matter energy and $H_{\rm G}$ the analogue 
quantity for the graviton state.
\par
It is now crucial that our present universe appears to be mostly 
driven by dark energy, and as such it is characterised by the Hubble 
radius,
\be
L
=
H^{-1}
\ ,
\label{Laa}
\ee
of the visible portion.
Furthermore, the presence of baryonic matter (stars and planets) 
defines a typical size $R_{\rm B}$, around which gravity is well 
approximated by Newtonian physics. These two length scales 
satisfy the hierarchy
\be
\Rh\ll R_{\rm B}\ll L
\ ,
\label{RhRbL}
\ee
where $\Rh= 2\,\gn\,m_{\rm B}$ is the Schwarzschild radius of a source 
of baryonic mass $m_{\rm B}$. 
The quantum state of gravity should entail such scales. In particular, 
we expect to identify different regimes of gravity for each scale from 
the way both the number of gravitons $\Ng$ and their typical energy 
$\eg$ scale with the mass $m=m(r)$ and the size $r$ of the region we 
are considering. 
\par
In the corpuscular description, one is mainly concerned with {\sl self-gravitating} 
systems, {\em i.e.}~compact sources of typical size $R_{\rm B}$.
To this class belong both marginally bound systems, which are described by
BEC at the critical point  (black holes) and non-marginally bound systems
(compact stars, horizonless objects) which are described by BEC away from 
the critical point. In terms of the graviton coupling $\alpha\simeq\lp^2/r^2$ 
the two regimes respectively correspond to $\alpha=1/\Ng$ and
$\alpha<1/\Ng$~\cite{Dvali:2011aa}. 
\par
In terms of the Hamiltonian constraint~\eqref{hc} the marginally bound 
condition corresponds to systems for which the mass is equal to the 
graviton interaction energy~\cite{Casadio:2016zpl}.
At small scales, {\em i.e.}~for $r$ of the order of the size of compact 
sources, the number of gravitons $\Ng$ affected by the presence of matter 
sources can be obtained by describing the Newtonian (and first 
post-Newtonian~\cite{Casadio:2016zpl,Casadio:2017cdv}) potential by means 
of a quantum coherent state, for which one generically 
finds a quadratic scaling of $\Ng$ with the mass~\cite{Dvali:2011aa},
\be
\Ng
\sim
\frac{m_{\rm B}^2}{\mpl^2}
\ ,
\label{qs}
\ee
where $m_{\rm B}$ is the mass of the localised baryonic source.
Since the (negative) Newtonian energy is given by
\be
U_{\rm N}
\simeq
\Ng\,\varepsilon_{\rm G}
\simeq
-\frac{\gn\,m_{\rm B}^2}{r}
\ ,
\label{Un}
\ee
the typical energy of the individual (virtual) quanta is again given by 
the Compton relation
\be
\varepsilon_{\rm G}
\simeq
-\frac{\lp\,\mpl}{r}
\ ,
\ee
and, using the mass/radius relation for black holes, $m_{\rm B}\simeq r=\Rh$,
Eq.~\eqref{Un} implies an holographic scaling with $r$, namely 
\be
\Ng
\sim
\frac{r^2}{\lp^2}
\sim
-\frac{1}{\rho_{\rm H}}
\ , 
\label{dSgravitons1} 
\ee
where $\rho_{\rm H}$ is the (negative) graviton energy density around a black hole.
\par
Notice that, for non-marginally bound gravitational systems, the scaling 
relation~\eqref{qs} still holds~\cite{Casadio:2016zpl, Casadio:2017cdv}, 
but the holographic~\eqref{dSgravitons1} does in general not. The corpuscular 
description can be generalized to cosmological space-times~\cite{Dvali:2013eja} 
in absence of baryonic matter. In this framework, the dS~universe of size $L$, 
sourced by a constant dark energy density $\rho_\Lambda$, can be described, 
similarly to a black hole, as a critical BEC~\cite{Binetruy:2012kx}.
In fact, the main feature of the dark energy sourcing the dS space-time, 
namely that it satisfies the vacuum equation of state $p=-\rho_\Lambda$,
is naturally realised in a BEC, as was shown in
Refs.~\cite{Volovik:2001fm,Volovik:2003fe,Volovik:2004gi,Nishiyama:2004ju}.
\par
An ideal universe of size $L$ solely containing self-coupled gravitons as 
a description of vacuum (dark) energy should behave like the de~Sitter 
space-time.
In GR, one then needs a cosmological constant term, or 
constant vacuum energy density $\rho_\Lambda$, so that the Friedman 
equation reads
\be
H^2\equiv
\left(\frac{\dot a}{a}\right)^2
\simeq
\gn\,\rho_\Lambda
\ .
\ee
Upon integrating on the volume inside the Hubble radius~\eqref{Laa}, we obtain
\be
L
\simeq
\gn\,L^3\,\rho_\Lambda
\simeq
\gn\,m_\Lambda
\ .
\label{LgM}
\ee
This relation looks like the expression of the horizon radius for a black 
hole of ADM mass $m_\Lambda$, which has led to conjecture that the dS 
space-time could likewise be described as a condensate of gravitons~\cite{Dvali:2013eja,Binetruy:2012kx}.
One can in fact introduce a corpuscular description on assuming that the 
(soft virtual) graviton self-interaction gives rise to a condensate of $N_\Lambda$ 
gravitons of typical Compton length equal to $L$~\cite{Dvali:2013eja}, 
so that the total (positive dark) energy
\be
m_\Lambda
\simeq
\NL\,\varepsilon_\Lambda
\simeq
N_\Lambda\,\frac{\lp\,\mpl}{L}
\ ,
\label{HL}
\ee
and, from Eq.~\eqref{LgM}, it follows immediately that
\be\label{dSgravitons4}
N_\Lambda
\sim
\frac{m_\Lambda^2}{\mpl^2}
=\frac{L^2}{\lp^2},
\ee
which shows that one needs a huge $N_\Lambda\gg 1$ for a macroscopic universe.
Note also that we have
\be
\rho_\Lambda
\sim
\frac{m_\Lambda}{L^3}
\sim
\frac{1}{N_\Lambda}
\ ,
\label{rhoL}
\ee 
so that the number of gravitons in the vacuum increases for smaller vacuum energy,
and
\be
L
\sim
m_\Lambda
\sim
\frac{1}{\sqrt{\rho_\Lambda}}
\ .
\ee
\subsection{Holographic regimes of gravity}
We have shown that black holes and the dS universe can be described by 
a critical BEC of gravitons we dubbed DEC.
We have also seen that criticality for the BEC implies the holographic 
scalings~\eqref{dSgravitons1} and~\eqref{dSgravitons4} for $\Ng$.
Being all the gravitons packed in the ground state, the entropy of the 
DEC is given by $\Ng$, implying that Eqs.~\eqref{dSgravitons1} 
and~\eqref{dSgravitons4} are equivalent to the BH area 
law~\footnote{Factors of order one will be usually neglected unless necessary.}.
\par
Eqs.~\eqref{qs} and \eqref{dSgravitons4} define the {\sl Holographic Regimes of gravity}:
for volumes of both cosmological (in absence of baryonic matter) and 
Newtonian size, one can argue that the relevant number of gravitons 
scales holographically, that is
\be
\Ng^A(r)
\sim
\frac{m^2(r)}{\mpl^2}
\sim
\frac{r^2}{\lp^2}
\ ,
\quad
{\rm for}\
r\simeq L
\
{\rm and}\
r\simeq R_B
\ ,
\label{Nga}
\ee
where $m=m(r)$ is an appropriate mass function inside the volume.
More precisely, $\Ng(L)$ can be viewed as the total number of 
gravitons inside the visible universe, whereas $\Ng(R_{\rm B})$ is the 
number of gravitons that respond locally to the presence of the baryonic
sources of mass $m_{\rm B}$, by changing their energy from $\eg(L)$ 
to some $\eg(R_{\rm B})<\eg(L)$ in order to enforce the Newtonian dynamics.
The holographic scaling~\eqref{Nga} therefore applies to two very 
different, albeit equally non-extensive, regimes of gravity. 
\par
The holographic scaling relations~\eqref{dSgravitons1} and~\eqref{qs} 
were first found for black holes~\cite{Dvali:2013eja}, and only 
Eq.~\eqref{qs} was then shown to hold for general compact 
sources in Refs.~\cite{Casadio:2016zpl,Casadio:2017cdv}.
From Eqs.~\eqref{qs}, it follows
that we get the BH area law~\eqref{Nga} in the regime where the
relevant mass $m=m(r)$ of the condensate scales linearly with the size $r$
of the source.
\par
The holographic regime of gravity holds for sure in the case of 
black holes and the de~Sitter space, and we assume that 
Eq.~\eqref{Nga} also remains a very good approximation at all 
typical scales $r$ for which gravity is well described by GR.
This assumption is based on the fact that the holographic 
nature of gravity is a generic consequence of the Einstein-Hilbert 
action.
Note, however, the change in sign of the graviton energy from 
the positive cosmological mass~\eqref{HL} to the negative Newtonian 
energy~\eqref{Un}: this is a clear signal that the two holographic regimes,
at small and very large scales, respectively, are indeed different, 
which suggests that at intermediate scales the behaviour of gravity 
deviates from the holographic description, as we will see in the next section.
\par
Before we proceed, a word of caution is in order:
since the gravitons in the condensate are considered as virtual 
(non-propagating) modes, their number $\Ng$ is not directly observable, 
nor is their individual energy $\eg$. In fact, one can think of these 
quantities as convenient intermediate variables which will not appear in 
our final expression for the matter dynamics.
These gravitons could however become observable if they are scattered off 
the coherent state, for instance by their self-interaction, which leads 
to the depletion of the DEC.
This effect produces the Hawking radiation around black holes~\cite{Dvali:2011aa}
and primordial perturbations during inflation~\cite{Dvali:2013eja,Casadio:2015xva},
but will be totally neglected in this work. 
\subsection{Extensive regime of gravity}
\label{ss1}
There are several reasons, coming both from the microscopic and from 
the emergent space-time description, for arguing that the holographic 
regime~\eqref{Nga} of gravity can not hold throughout the whole range of 
scales~\eqref{RhRbL}.
In particular, this implies the existence of a new infrared scale $\Rh<r_0<L$,
where the behaviour of gravity deviates from the holographic description. 
\par
The first indication comes from the fact that the two holographic 
regimes at small and very large scales, although satisfying the same 
scaling relation~\eqref{Nga}, are indeed different.
We recalled above that the graviton energy changes in sign going from
the positive cosmological mass~\eqref{HL} to the negative Newtonian
energy~\eqref{Un}.
This implies the two holographic regimes must be connected by a {\sl mesoscopic\/}
phase, in which gravity may deviate from the holographic behaviour~\eqref{Nga}.  
\par
The second indication comes from Verlinde's argument about the 
pattern of entanglement entropy in dS~space~\cite{Verlinde:2016toy}. 
Unlike black holes, the dS~space-time must contain a thermal 
volume contribution to the entanglement entropy, coming from very low 
energy modes.
In our description of the dS~space-time, this implies an extensive term for the
graviton number associated with the DEC.
\par
The third and strongest indication comes from the fact that, locally, 
without baryonic matter, the DEC of the dS~space-time has a constant 
energy density characterized by an extensive behaviour.
In fact, at galactic scales, we cannot consider the cosmic condensate 
as a whole, but just as a medium with (positive) constant energy 
density $\rho_{\rm G}$ equal to the cosmological value~\eqref{rhoL}, 
that is
\be
\rho_{\rm G}
\simeq
\rho_\Lambda
\sim
\frac{\mpl}{L^2\,\lp}
\ .
\ee
The total graviton energy inside a region of size $r$ is therefore given by
\be 
m_{\rm G}(r)
\simeq
\frac{4\,\pi}{3}\,\rho_\Lambda\,r^3
\sim
\frac{\mpl\,r^3}{L^2\,\lp}
=
\Ng(r)\,\varepsilon_\Lambda
\ .
\ee 
The number of gravitons contained in this spherical region is therefore 
an {\sl extensive} quantity, scaling as the volume, 
\be
\Ng
\simeq
\frac{r^3}{L\,\lp^2}
\sim
\frac{m_{\rm G}(r)}{\mpl}
\label{dSgravitons5}
\ee
where we again assumed the Compton relation $\varepsilon_\Lambda\sim \lp\,\mpl/L$
from Eq.~\eqref{HL}. 
A crucial check for the validity of the scaling relation~\eqref{dSgravitons5} 
is that it correctly reproduces the cosmological relation~\eqref{dSgravitons4}
precisely for $r=L$.
\par
We are therefore lead to assume that, if baryonic matter is 
totally neglected, at the intermediate scales $R_{\rm B}\ll r\ll L$, 
the graviton state is approximately described by the extensive 
regime~\eqref{dSgravitons5}, {\em i.e.}~it is ruled by the 
{\sl Extensive Regime of gravity}:
\be
\Ng^V(r)
\sim
\frac{m}{\mpl}
\sim
\frac{r^3}{L\,\lp^2}
\ ,
\quad
{\rm for}\
R_{\rm B}\ll r\ll L
\ .
\label{Ngv}
\ee
This behaviour will be argued to interpolate somehow between the two 
(different) holographic regimes~\eqref{Nga} at $r\simeq L$ and $r\simeq R_{\rm B}$.
One of the main results we will present here is that it is the tension 
between the two scalings~\eqref{Nga} and \eqref{Ngv} that leads to 
deviations from the local Newtonian dynamics~\cite{Verlinde:2016toy}:
the response of the graviton condensate to the presence of baryonic matter 
makes both the holographic and the extensive regimes important at galactic scales.  
\par
The physical picture behind this corpuscular description is again similar to 
Verlinde's~\cite{Verlinde:2016toy}.
For compact sources of size  $R_{\rm B}\simeq \Rh$ and at cosmological scales $L$,
gravity allows for a corpuscular description in which it is described by a critical 
BEC of gravitons.
The effective theory in these two regimes is GR~\footnote{Since the universe
is expanding, one might argue that the cosmological description is in fact closer to 
a modified $f(R)\simeq R^2$ theory of gravity~\cite{Casadio:2017cdv}.},
whose peculiar non-extensive, holographic character is encoded by the 
relations~\eqref{dSgravitons1} and~\eqref{dSgravitons4}. 
Notice that these two regimes corresponds to length scales differing by 
several orders of magnitude (about $60$ if we take $r=\lp$ and $r=L$), 
and the same holds for the graviton wavelengths in the two regimes. 
\par
A specific merit of the corpuscular picture we started to build is however 
that these two holographic regimes are truly different, as the 
relation~\eqref{Nga} refers to the total number of gravitons in the 
cosmological condensate for $r\simeq L$, whereas it only counts 
the number of gravitons affected by the local matter sources for 
$r\simeq R_{\rm B}\gtrsim \Rh$.
We recall once more the difference is clearly signalled by the opposite signs
of $\varepsilon_\Lambda>0$ and $\varepsilon_{\rm B}<0$.  
\par
At intermediate scales the condensate has the intrinsic extensive behaviour 
\eqref{dSgravitons5}, which is a peculiar feature of thermalization processes 
(corresponding to the slow dynamics of glassy systems in Verlinde's description).
Strictly speaking the graviton number $\Ng^V$ inside a spherical region is
not physically measurable. In fact $\Ng^V$ is not conserved and, for a small 
region, it is expected to have large relative fluctuations. For regions of 
galactic or cosmological size, the relative fluctuations are small but we can 
hardly conceive a physical process apt to measure $\Ng^V$.
On the other hand, our final results are independent from $\Ng^V$ and we do 
not need to be concerned about its measurability. 
\par
In principle, there could be concerns about the impact that an extensive,
volume-scaling, term for $N_{\rm G}$ can have on the cosmological evolution, 
in particular for late-time cosmology.
At late times, the cosmological dynamics is described by the holographic regime
characteristic of the dS space-time as discussed above. 
Actually, it has been recently shown by Carroll~{\it et al}~\cite{Carroll:2017kjo}
that this is a quite general result.
Assuming the validity of a generalized second law of thermodynamics and that
the entropy increases up to a finite maximum value, any Robertson-Walker
space-time must approach a dS space-time in the future, independently of the
gravitational dynamics and matter content of the universe. 
In their argument, Carroll~{\it et al}~assume the presence of a constant density  
term in the generalized entropy, which has the same form of our extensive 
term~\eqref{dSgravitons5}.
However, they show that at late times, {\em i.e.}~for large values of the scale factor,
this  term is  subleading with respect to the holographic one, the latter approaching
a constant value and scaling like the area of the dS horizon.
Translated in our corpuscular description, this means that our extensive
term~\eqref{dSgravitons5} plays a role at intermediate galactic scales,
but becomes completely irrelevant for the late-time cosmological evolution.  
\subsection{Baryonic matter and the emergence of a dark force}
So far we have considered the cosmological condensate without
baryonic matter. 
One could just consider baryonic matter always existed
inside the DEC, initially in a very diluted form, so that its effect on the 
gravitons of the cosmological BEC was initially negligible.
In time, the baryonic matter clumped and started affecting the DEC locally,
which is the situation we find in the universe today.
In particular, the presence of local baryonic sources pulls out 
gravitons from the DEC, which give rise to the local gravitational
forces.
Alternatively, the simplest way to introduce baryonic matter in our scenario is to 
assume that it arises as bound states in the DEC, {\em i.e.}~to consider
it as produced by gravitons pulled out from the cosmological 
condensate at the typical matter scales $R_{\mu}$, where $\mu$ denotes 
the mass of single point-like matter sources.
This may occur owing to density perturbations in the BEC.
An uniform, spherically symmetric over-density region of the BEC is isotropically
compressed, because pressure gradients act only on the surface of the sphere,
generating a compact source of baryonic matter, which can itself be described by a 
non-critical BEC or by critical BEC if the critical density is reached 
and a black hole is formed.
\par 
In the next sections, we will first discuss the behaviour of the 
condensate with baryonic matter in the diluted approximation, when we 
can neglect the local reaction of the condensate. When we go beyond 
this approximation, we have to take into account the reaction of the 
cosmic condensate to the presence of the baryonic matter.
We will see that this can be described as a dark force, mediated by 
gravitons pulled out from the cosmic BEC at galactic scales, which can 
explain the phenomenology at galactic scales commonly attributed to dark 
matter.
\par
An important point to be stressed is that the dark force is a local effect. 
The cosmological BEC at horizon scales $L$ remains largely unaffected.
This means that deviations from Eq.~\eqref{dSgravitons4} for the cosmological BEC 
remain negligible at present, albeit they are crucial in order to describe 
the local dynamics properly, as we are going to start showing next.
Our description is consistent as long as we are only concerned with the 
gravitational dynamics at galactic scales and we do not use our model to 
describe the whole cosmological history of our universe. In order to do 
this, it is likely that more input is needed. 
\section{Baryonic Matter in the diluted approximation}
\label{sec3}%
\setcounter{equation}{0}
We now want to see in more details what happens when very diluted 
baryonic matter is formed on top of the condensate of gravitons. 
In this approximation, matter can be considered as being made of, 
say $N_\mu$ almost point-like sources of mass $\mu$, at rest and 
equally distanced very far apart. We can therefore neglect the local 
reaction of the condensate to their presence, which also means that 
the gravitational interactions among matter sources are negligible.
Since sources are homogeneously distributed, our results should also 
be a good approximation for baryonic matter with homogeneous density.
We will see that the leading-order effect of baryonic matter is to 
subtract gravitons from the condensate.
\subsection{Diluted matter in the de~Sitter universe}
Let us first see what happens when we introduce baryonic matter 
into the de~Sitter universe, whose metric takes the form
\be
\d s^2
=
-f(r)\,\d t^2
+f^{-1}(r)\,\d r^2
+r^2\,\d\Omega^2
\ .
\label{dS}
\ee
Since in the diluted approximation the cumulative effect of many 
sources is just the sum of the single contributions, we start by 
considering the case of a single point-like source of mass $\mu$.
In the weak field regime, the metric function in Eq.~\eqref{dS} is 
given by the Schwarzschild-dS form
\be
f(r)
=
1-\frac{r^2}{L^2}+2\,\phi(r)
\ ,
\ee
where 
\be
\phi(r)
=
-\frac{\gn\,\mu}{r}
\ee
is the Newtonian potential generated by the source of mass $\mu$. 
The size $L_H$ of the cosmological horizon can be found by solving 
the condition $f(r)=0$ for small departures from $L$ 
(i.e.~for $|\phi|\ll 1$), which yields
\be
L_H
=
L\left[1+\phi(L)\right]
+o(\phi^2)
\sim
L- \lp\,\mu/\mpl
\ .
\ee
Adding $N_\mu$ similar matter sources would reduce the Hubble radius to
\be
L_H
\sim
L- \frac{1}{2}\, N_\mu\,\Rh
\ ,
\label{rh}
\ee
where here $\Rh= 2\,\gn\,\mu$ is the typical gravitational radius of each source.
The effect of the presence of diluted matter is thus to reduce the size of the 
cosmological horizon, which in turn implies a number of gravitons 
in the cosmological condensate $\Ng<N_\Lambda$ according to 
Eq.~\eqref{dSgravitons4}.
Let us note, however, that such a change is relatively minuscule because
of the hierarchy~\eqref{RhRbL}, and the fact that baryonic matter accounts
for at most $5\%$ of the total energy in the universe.
We can therefore safely neglect the difference between $L_H$ and $L$
in the following.
\subsection{Diluted matter in the corpuscular model}
Before we introduce the diluted baryonic matter in the DEC, let 
us refine the corpuscular description of the dS universe.
In Refs.~\cite{Casadio:2016zpl, Casadio:2017cdv}, it was shown that 
the maximal packing condition which yields the scaling relations~\eqref{Nga} 
for a black hole actually follow from the energy balance~\eqref{H0} 
when matter becomes totally negligible. In the present case, matter 
is absent {\em a priori\/} and $H_{\rm B}=0$, so that one is left with
\be
H_{\rm G}^{(0)}
=
U_{\rm N}^{(0)}
+
U_{\rm PN}^{(0)}
=
0
\ ,
\label{H00}
\ee
with the negative Newtonian energy
\be
U_{\rm N}^{(0)}
\simeq
N_\Lambda\,\varepsilon_{\Lambda}
=
-N_\Lambda\,\frac{\lp\,\mpl}{L}
\ ,
\label{UN0}
\ee
and the positive ``post-Newtonian'' contribution
\be
U_{\rm PN}^{(0)}
=
N_\Lambda\,\frac{\sqrt{N_\Lambda}\,\lp^2\,\mpl}{L^2}
\ .
\label{UPN0}
\ee
One therefore recovers the scaling relation~\eqref{dSgravitons4} 
from Eq.~\eqref{H00} with no extra ``vacuum energy''~\cite{Casadio:2017cdv}.
\par
The same result~\eqref{rh} can now be obtained using the Hamiltonian 
constraint~\eqref{H0} in which we include the contribution of $N_\mu$ 
diluted baryonic sources of mass $\mu$,
\be
H_{\rm B}^{(1)}
=
N_\mu\,\mu
\ .
\ee
Since matter is very diluted and cold, $\mu$ again just equals the 
proper mass, and local gravitational energy is negligible.
We can therefore write
\be
\label{eb}
H_{\rm G}^{(1)}
=
U_{\rm N}^{(1)}
+
U_{\rm PN}^{(1)}
\ ,
\ee
where the Newtonian and post-Newtonian terms have the forms given in
Eqs.~\eqref{UN0} and~\eqref{UPN0}.
The energy balance~\eqref{H0} then tells us that the condensate  
must respond to the presence of this homogeneous matter by changing the 
graviton number $\NL$, that is
\be
\NL\,\frac{\lp\,\mpl}{L}
\simeq
N_\mu\,\mu
+
\NL^{3/2}\,\frac{\lp^2\,\mpl}{L^2}
\ ,
\label{H1}
\ee
which yields
\be
L
&\!\!=\!\!&
\frac{\lp\,\mpl\,\NL}{2\,N_\mu\,\mu}
\left(
1-\sqrt{1-\frac{4\,N_\mu\,\mu}{\sqrt{\NL}\,\mpl}}
\right)
\nonumber
\\
&\!\!\simeq\!\!&
\sqrt{\NL}\,\lp
+
N_\mu\,\lp\,\mu/\mpl
+
\mathcal{O}(\NL^{-1})
\ ,
\label{LM}
\ee
where we used $N_\mu\,\mu\ll\NL\,\mpl$ for our dark energy dominated universe.
Using now the scaling~\eqref{dSgravitons4}, {\em i.e.}~$N_\Lambda\sim L_H^2/\lp^2$,
one easily recovers Eq.~\eqref{rh}.
\par
The fact that the two estimates, (respectively based on the form of the 
Schwarzschild-dS metric and on the corpuscular model, give the same 
result for the change of the Hubble horizon due to presence of baryonic 
matter is a highly non trivial check of the validity of our BEC 
description of the dS~universe, and in particular  of the validity of
the energy balance~\eqref{eb} and of the form of the post-Newtonian 
term $U_{\rm PN}$.
\subsection{Diluted matter and scalings of the graviton number}
The change in the dS horizon size~\eqref{rh} 
induced by the baryonic matter will result in a reduction of the 
number of gravitons $\NL$ with energy $\varepsilon_\Lambda$
according to Eq.~\eqref{dSgravitons4}, that is
\be
\delta \NL
\simeq
-\frac{2\,\mu\,L}{\mpl\,\lp}
\ ,
\ee
where $N_\Lambda$ is given  in~\eqref{dSgravitons4}). 
The same result holds also for a black hole of mass $\mu$, with $L$ 
replaced with the black hole radius $\Rh$~\cite{Verlinde:2016toy}.
Actually, this result is a quite generic consequence of the holographic 
scaling~\eqref{Nga} for the graviton number.
In fact, let us take a sphere of radius $r\ll L$, for which the number 
of gravitons in the condensate inside this sphere is given by 
Eq.~\eqref{Nga}, and compare the change of the graviton number as a 
function of the radial distance from the centre of the sphere with and 
without matter. Without the mass, the radial distance $s$ is equal to 
$r$, whereas a baryonic point-like mass $\mu$ at the center of the 
sphere changes the radial distance of a quantity equal to 
$\d s\simeq[1-\phi(r)]\,\d r$, according to the weak field limit of 
the Schwarzschild metric.
Thus, the number of gravitons in Eq.~\eqref{Nga} changes due to the
presence of matter according to
\be
\label{kly}
\frac{\d(\delta \NL)}{\d s}
=
\frac{\d}{\d s}
\left(\left.\NL\right|_{\mu\neq0}
-\left.\NL\right|_{\mu=0}\right)
\simeq
\phi(r)\,\frac{\d\NL}{\d r}
\simeq
-\frac{2\,\mu}{\mpl\,\lp}
\ .
\ee
On the other hand, in the diluted approximation $|\phi(r)|\ll1$ 
and $\d r\simeq\d s$. We can thus write the previous equation as 
\be
\frac{\d(\delta N_\Lambda)}{\d r}
\simeq
-\frac{2\,\mu}{\mpl\,\lp}
\ .
\ee
For future convenience, we will define $\Nb=-\delta N_\Lambda$ as 
the number of gravitons subtracted from the cosmological condensate 
(the DEC) inside a sphere of radius $r$ by the presence of 
the baryonic source of mass $\mu$. By integrating the above 
equation, one finds
\be
\Nb
\simeq
-\frac{2\,\mu\,r}{\mpl\,\lp}
\ .
\label{Nb}
\ee
Extending the validity of Eq.~\eqref{rh} from the cosmological 
horizon to a region of radius $r$ as given in Eq.~\eqref{kly} is 
a quite strong and highly non trivial assumption.
In the corpuscular description of gravity, this implies that we are assuming
not only the whole dS space filled with dark energy can be considered as a 
graviton condensate with Compton length $L$, but that this description 
also holds for regions of any size $r$, and for those gravitons with Compton 
length of order $r$.
The rational behind this assumption is the fact that 
at solar system scales we know that gravity is well described by GR,
whose action is directly related with the holographic 
scaling~\eqref{Nga}.
It should be stressed that this holds only in the holographic 
regime of gravity~\eqref{Nga} but not in the extensive regime~\eqref{Ngv}.
This means that our universe looks like a critical graviton condensate at
small (solar system) scales and very large (Hubble radius) scales,
whereas at intermediate (galactic) scales we see an extensive 
behaviour.
In order to give a precise meaning for the transition from 
cosmological to intermediate scales, we can use arguments similar to those 
used by Verlinde in Ref.~\cite{Verlinde:2016toy}.
\par 
We suppose that, as shown in \cite{Verlinde:2016toy}, the ``dark matter''
effects arise from the competition between the ``area-law''~\eqref{Nga} 
and volume behaviour~\eqref{Ngv} for the graviton number. 
This implies the existence of two regimes: the baryonic matter 
dominated regime in which $\Nb(r)> \Ng^V(r)$ and a dark energy dominated regime 
$\Nb(r)< \Ng^V(r)$. In particular, we expect the dark force effects to be 
negligible for  $\Nb(r)\gg \Ng^V(r)$. 
Let us now look for the transition between these two regimes, when 
the corresponding graviton numbers become comparable, that is 
$|\Nb(r)|\simeq \Ng^V(r)$, or
\be
\frac{2\,\mu\,r}{\mpl\,\lp}
\simeq
\frac{r^3}{\lp^2\,L}
\ .
\label{area/volume_regime}
\ee
When this equality holds, most of the dark energy gravitons in 
the cosmological condensate contained inside the volume of size 
$r$ are affected by the presence of the source of mass $\mu$,
and we obtain
\be
r
\equiv
r_0
\simeq
\sqrt{\frac{2\,\mu}{\mpl}\,L\,\lp}
=
\sqrt{\Rh\,L}
\ ,
\label{area/volume_scale}
\ee
where $r_0$ is the mesoscopic scale introduced in Section~\ref{ss1}.
For a given (spherical) region with a certain 
amount of mass $\mu$ localised about its center, $r_0$ sets the 
scale at which dark matter phenomena are not negligible. 
Using for $\mu$ the value for the mass of a galaxy in Eq.~\eqref{area/volume_regime},
one finds the observationally correct order of magnitude for deviations from
the Newtonian dynamics.
For instance, for a typical spiral galaxy with $m_{\rm B}= 10^{11}$ solar masses,
we have $r_0= 6\,$kpc, whereas for a typical dwarf galaxy with $m_{\rm B}= 10^{7}$ 
solar masses, we have $r_0= 80\,$pc. 
\par
To describe the transition between the holographic and the extensive 
regimes, it is convenient to introduce the local (size-dependent) parameter
\be
\gamma
=
\frac{\Nb}{\Ng^V}
\ . 
\label{volumeArea}
\ee
For $\gamma>1$, we are in the area-scaling regime~\eqref{Nga}, 
where baryonic matter dominates, gravity is well described by GR,
and most of the gravitons in the fluid belong to the condensate.
Conversely, for $\gamma<1$, we are in volume-scaling 
regime~\eqref{Nga} and dark energy dominates. 
In this regime, the effects of the dark energy gravitons on baryonic 
matter are not negligible and give rise to the dark matter phenomena. 
\par
Let us conclude with some comments about the physical meaning of the 
diluted approximation and on the meaning of Eq.~\eqref{volumeArea}. 
Within this approximation, baryonic matter has no local 
gravitational interactions with the condensate.
On the other hand, it also has no effects at cosmological scales.
Rephrased in terms of the graviton number, the diluted regime applies 
in the region where $\gamma>1$, {\em i.e.}~when most gravitons inside 
a sphere of radius $r$ belong to the {\sl local} condensate 
({\em i.e.}~we are considering the sphere of radius $r$ as a condensate 
of gravitons of Compton length $r$).    
\section{Clumped matter and emergence of the dark force}
\label{sec4}
\setcounter{equation}{0}
Let us now describe what happens when we go beyond the diluted approximation 
and baryonic matter begins to clump.
The $N_\mu$ point-like sources of  mass $\mu$ form clusters of baryonic matter
with typical mass $m_{\rm B}(r)=N_\mu\, \mu$.
For simplicity, we consider a mass distribution with spherical symmetry.
Now the DEC will react to the presence of matter, and we will interpret this reaction
as a dark force responsible for the phenomenology commonly attributed to dark matter 
which reproduces correctly the MOND acceleration.
\par
We first assume that only a fraction of the gravitons in the DEC are affected by the local matter,
so that the condensate reacts not at the full cosmological scale $L$, but at a local scale of size $r$.
In particular, since we are considering spherically symmetric sources, the baryonic matter
of mass $m_{\rm B}$ will pull the gravitons out of the DEC from inside the sphere of radius $r$
with a dark energy mass given by $M=M(r)$.
Therefore we now have three scales in our problem:
the typical size of the matter lumps $\Rb$, the range of the condensate reaction $r$
and $L$, which satisfy the hierarchy 
\be
\Rh\ll \Rb\lesssim r \ll L
\ .
\label{hierarchy}
\ee
\par
In the following we will consider the dynamics of test particles at 
distances $r\gg R_{\rm B}$, so that the baryonic source of mass $m_{\rm B}(r)$
can be well approximated by a point-like source.
Physically, this means that we are considering the dynamics of galaxies
at distances far away from the galactic core.
We will first briefly review the results of Ref.~\cite{Cadoni:2017evg} 
based on a balance between the number of gravitons, we will then give the 
description based on the Hamiltonian constraint, finally we will proceed 
by using the competition between the area and volume regimes to derive the 
``dark acceleration''.
\subsection{ Matter clumping and  graviton number balance}
\label{ss:cdf}
The starting point of the analysis in Ref.~\cite{Cadoni:2017evg} 
is that,  in a corpuscular description of gravity, the gravitational acceleration
felt by the test particle is the macroscopic manifestation of the self-interaction
of gravitons in the condensate.
It can therefore be expressed in terms of their Compton energy $\varepsilon$
and specific number $\Neff$ of gravitons involved in the process as
\begin{equation}
a(r)  
\simeq
\frac{\varepsilon^2(r)}{\mpl^2\,\lp}\, \sqrt{\Neff}
\ . 
\label{aNewt}
\end{equation}
Moreover, this {\sl corpuscular acceleration\/} formula holds for both 
condensed and non-condensed gravitons.
Consider now the reaction of the cosmological BEC of total mass $m_{\Lambda}$
to the presence of the baryonic matter source of mass $m_{\rm B}(r)$.
Since $m_{\rm B}\ll m_{\Lambda}$, most of the gravitons will remain in the
condensed phase and their number is given, according to Eq.~\eqref{Nga}, by 
\begin{equation}
\label{N0}
	N_\DE
	\sim
	\frac{\left(m_{\Lambda}-m_B\right)^2}{\mpl^2}
\end{equation}
On the other hand, the total number of gravitons in the system is given by 
$N_{\Lambda}\sim{m_{\Lambda}^2}/{\mpl^2}$.
This implies that there are $N_{\Lambda}-N_\DE$ gravitons which are not in
the condensed phase and, therefore, behave differently from the condensate.
Since the number of gravitons
which give rise to the local gravitational potential generated by the baryonic mass
is $N_{\rm B}={m_B^2}/{\mpl^2}$ and, from Eqs.~\eqref{N0}, we have 
\be
N_{\Lambda}-N_\DE\sim \frac{L\,\mb}{\lp\, \mpl} - \frac{m_B^2}{\mpl^2}
\ ,
\ee
it follows that there are $N_\DF\sim	{L\, \mb}/{\lp\,\mpl}$ gravitons which mediate
the interaction between the baryonic matter and the DEC.
\par
The effective number of non-condensed gravitons $N_{\DF}(r)$
that contribute to the acceleration of a test particle at the radius $r$ can be
guessed by requiring that its overall scaling is again holographic and must
depend on the baryonic mass $m_{\rm B}$.
This yields
\begin{equation}
\label{NDF.r}
	N_{\DF}(r)
	\sim
	\frac{r^2\, m_{\rm B}(r)}{\lp\,\mpl\, L}
	\ .
\end{equation}   
From Eqs.~\eqref{aNewt} and~\eqref{NDF.r} with $\Neff=N_{\DF}(r)$, we obtain 
\begin{equation}
\label{aDF.N}
	|a_\DF(r)| \sim \sqrt{\frac{\gn\,m_{\rm B}(r)}{L\,r^2}}
	\sim
	\sqrt{\frac{\ab(r)}{L}}
	\ , 
\end{equation}
which is the MOND acceleration~\eqref{MOND} up to a numerical factor.
\subsection{Matter clumping and energy balance}
\label{ss:eb}
In this section we will alternatively derive the MOND acceleration~\eqref{MOND} 
using the Hamiltonian constraint~\eqref{hc}.
\par
Once the regular matter starts clumping, the matter energy changes to 
\be
H_{\rm B}
=
m_{\rm B}
+
E_{\rm B}
\ ,
\label{HM>}
\ee
where $m_{\rm B}\simeq N_\mu\,\mu$ and $E_{\rm B}$ accounts for the 
total kinetic energy of matter and non-gravitational interactions. 
Some gravitons will acquire a new Compton length in response to the 
local lumps of matter, and the gravitational Hamiltonian in the constraint Eq.~\eqref{hc} 
takes the form
\be
H_{\rm G}
=
H_{\Lambda}
+
H_{\rm BG}
+
H_{\rm DF}
\ ,
\label{Hgbd}
\ee
where $H_{\Lambda}$ is the energy of the DEC, whose 
specific form is not essential for the present derivation;
$H_{\rm BG}$ is the Newtonian gravitational energy of the 
localised matter sources,
\be
H_{\rm BG}
=
-\frac{\gn\,\mb^2}{\Rb}
=
-\Nb\,\frac{\lp\,\mpl}{\Rb}
\ ,
\label{Hbg4.1}
\ee
with $\Nb$ the number of soft gravitons whose Compton length equals 
the typical size $\Rb$ of matter lumps~\footnote{There would also be a (positive)
post-Newtonian energy but we shall neglect that as it is much smaller than
$H_{\rm BG}$ for compact sources far from becoming black holes.};
finally, the ``dark force'' term is given by the gravitational interaction 
energy between baryonic matter and dark energy of mass $M(r)$ inside the sphere 
of radius $r$, that is
\be
H_{\rm DF}
=
-\frac{\gn\,\mb\,M(r)}{r}
\ .
\label{Hdm}
\ee
\par 
We can rewrite $H_{\rm DF}$ in terms of an effective dark force mass $m_{\rm DF}$
as 
\be
\frac{\gn\,\mb\,M(r)}{r}
\simeq
\frac{\gn\,m_{\rm DF}^2}{r}
\ ,
\label{mDM1}
\ee
which implies the simple relation between masses
\be
\label{masses} 
m_{\rm DF}^2
=
m_{\rm B}\,M(r)
\ .
\ee
Because the dark matter term arises from the interaction of the baryonic 
source with the gravitons in the DEC inside the volume of size $r$,
the  energy of the gravitons will change to $\varepsilon\simeq \mpl\,\lp/r$.
From the extensive scaling~\eqref{Ngv}, it follows that  
\be
M(r)
\simeq
\frac{\mpl\,r^2}{\lp\,L}
\ 
\label{Mlambda2}.
\ee
\par
We can now evaluate the gravitational acceleration associated to the
dark force component~\eqref{mDM1} of the condensate. 
Using the estimate~\eqref{Mlambda2} in Eq.~\eqref{masses}, 
we obtain the dark acceleration
\be
a_{\rm DF}
\sim
\frac{\gn\,m_{\rm DM}}{r^2}
\simeq
\sqrt{\frac{1}{L}\,\frac{\gn\,m_{\rm B}}{r^2}}
=
\sqrt{\frac{a_{\rm B}(r)}{L}}
\ ,
\label{adf}
\ee
where $a_{\rm B}(r)$ is the Newtonian baryonic acceleration~\eqref{aN} at distances
$r$.
Again, this result indeed matches the MOND formula~\eqref{MOND} up to a
factor of $1/6$.
\par
We can further show that the above derivation, based on the energy balance~\eqref{Hgbd},
is perfectly compatible and consistent with the derivation in Section~\ref{ss:cdf},
which is instead based on the graviton numbers.
In fact, we can associate to the dark energy mass $M(r)$ interacting with the baryonic mass
a number of gravitons equal to the number of gravitons $N_{\rm DF}(r)$ pulled out from the DEC.
This number scales holographically as 
\be
\label{holoNDF}
N_{\rm DF}(r)
=
\frac{m_{\rm DF}^2}{m_p^2}
\ .
\ee
By combining Eqs.~\eqref{masses} and \eqref{holoNDF}, we find 
\be
\label{MDE}
M(r)
=
N_{\rm DF}(r)\,\frac{\mpl^2}{m_{\rm B}}
\ ,
\ee
and Eq.~\eqref{Mlambda2} finally yields the total 
number of gravitons associated to the dark force
\be
\label{NDM}
N_{\rm DF}(r)
=\frac{m_{\rm B}\, r^2}{\lp\,\mpl\,L}
\ ,
\ee
which exactly matches Eq.~\eqref{NDF.r} obtained in Section~\ref{ss:cdf}. 
\par
We further note the dark acceleration can also be written as a function of the 
number of dark gravitons, thus obtaining the same expression~\eqref{aNewt} 
found in in Section~\ref{ss:cdf}.
In fact, by combining Eqs.~\eqref{MDE} and~\eqref{masses}, we find 
\be
\label{darkacceleration}
a_{\DF}
=
\frac{\gn\,m_{\DF}}{r^2}
=
\frac{\lp}{r^2}\,\sqrt{N_{\DF}(r)}
\ ,
\ee
or, equivalently, using the Compton energy of the dark gravitons $\varepsilon=\mpl\,\lp/r$, 
\be
a_{\DF}
=
\frac{\varepsilon^2(r)}{\mpl^2\,\lp}\,\sqrt{N_{\DF}(r)}
\ ,
\ee
which is exactly the corpuscular acceleration~\eqref{aNewt} first introduced in
Ref.~\cite{Cadoni:2017evg} for $\Neff=N_{\DF}(r)$. 
\subsection{Area/volume competition and heuristic derivation of MOND}

In this section, we present a heuristic derivation of the MOND acceleration~\eqref{MOND},
which uses the Hamiltonian constraint~\eqref{Hgbd} and the competition between
the holographic and extensive regimes described in Section~\ref{sec2}.
The novelty is that we will be able to reproduce correctly also the numerical 
factors of Eq.~\eqref{MOND} in this scenario.
The key observation is that, owing to the fact that the DEC responds only locally
to the presence of baryonic matter, we can simply write the contribution $H_{\DF}$
in Eq.~\eqref{Hgbd} in terms of the energy subtracted from dark energy gravitons 
to generate the local Newtonian gravity.
\par
For simplicity, we consider baryonic matter in the form of a single point-like 
source of mass $\mb$, but the results can be easily generalised to the case of an 
extended but localised source inside a volume of size $\Rb$.
By analogy with the electromagnetic force, the energy density $\rho_{\rm G}$ associated 
with a gravitational (acceleration) field~\eqref{aN}, that is
\be
a_{\rm B}=-\frac{\gn\,\mb}{r^2}
\ ,
\label{aNN}
\ee
inside a sphere of radius $r$, and volume $V(r)={4\,\pi\,r^3}/{3}$, is given by 
\be
\rho_{\rm G}
=
\frac{a_B^2}{8\,\pi\,\gn}
\ .
\ee
where $\mb$ is the source of the gravitational field.
It is easy to find that the energy subtracted from dark energy gravitons in order to clump
the amount of matter $\mb$ inside the spherical region is, therefore,
\be
\label{energysubtracted}
E_{\rm G}
=
-\rho_{\rm G}\,V
=
-\frac{\gn\,\mb^2}{6\,r}. 
\ee
where $\mb$ now denotes the  baryonic mass contained inside $V(r)$.
\par
Consistently with Eq.~\eqref{masses}, we can view this energy as due 
to the existence of a ``dark force'', whose effective source is a
``dark mass'' $m_{\DF}$, which does work on the system.
In analogy to what happens at cosmological scales, we can think that the 
effect of the mass $\mb$ centred inside a spherical region of volume $V(r)$,
is to deform the sphere by an amount given by Eq.~\eqref{rh} with $L$ replaced by $r$.
The deformation is therefore,  
\begin{equation}
u(r)
=
\phi_{\rm B}(r)\,L
\ ,
\end{equation}
where $\phi_{\rm B}$ is the gravitational potential generated by the mass 
$\mb $ and $L$ is still the dS radius.
The work done by the ``dark force'' on the system will be given by
\be
\label{work}
W
=
F_{\DF}\,u(r)
=\frac{\gn^2\,m_{\DF}^2\,\mb}{r^3}\,L
\ .
\ee
It should be stressed that this contribution is of holographic nature:
it is the work done by the dark force to deforme the {\em surface\/} of 
the sphere. For energy conservation, it must equal the energy $E_{\rm G}$ 
contained in the volume $V(r)$.
By equating Eqs.~\eqref{energysubtracted} and~\eqref{work}, we easily obtain
\be
\label{MOND1}
\frac{\gn\,m_{\DF}^2}{r^2}
=
-\frac{m_{\rm B}}{6\,L}
\ .
\ee
If we now use the form for  the ``dark gravitational acceleration'' used in~\eqref{adf}, 
\be
a_{\DF}
=
-\frac{\gn\,m_{\DF}}{r^2}
\ 
\ee
and the Newtonian acceleration~\eqref{aNN},  Eq.~\eqref{MOND1} can be written as 
\be
\label{MONDacceleration}
a_{\DF}(r)
=
\sqrt{\frac{a_{\rm B}}{6\,L}}
\ ,
\ee
which exactly matches the MOND acceleration~\eqref{MOND}. 
Let us stress that Eq.~\eqref{MOND} is precisely obtained by identifying 
the {\sl volume\/} (extensive) subtraction~\eqref{energysubtracted} from 
the condensate with the dark {\sl area\/} (holographic) contribution~\eqref{work}.
\subsection{Emergent metric theory}
A key issue for every model of emergent gravity is the existence of an effective 
description reproducing Einstein's GR or at least a metric 
theory of gravity.
One must envisage the way in which the metric space-time structure of gravity
encoded in GR emerges out of the microscopic description.  
This is a quite stringent requirement and it is not enough to predict an infrared
modification of the laws of gravity, such as the MOND relation~\eqref{MOND}.
This relation must be embedded in the framework of GR or, at least,
in a metric theory of gravity describing a modification thereof.
This is for instance a drawback of Verlinde's original proposal~\cite{Verlinde:2016toy}.
The proposed modification of the laws of gravity at galactic scales reproduces the MOND
relation~\eqref{MOND}, but a metric covariant description of the model has not been
proposed yet (see, however, Refs.~\cite{Hossenfelder:2017eoh,Dai:2017guq,Cai:2017asf}).   
\par
The description of the emergent laws of gravity at galactic scales based on 
the BEC of gravitons proposed in this paper allows for an effective covariant 
metric description, which has the form of GR sourced by an anisotropic fluid. 
This can be done along the lines of Refs.~\cite{Faber:2005xc,Cadoni:2017evg}. 
We know that a dark energy dominated universe, {\em i.e.}~the DEC of gravitons,
can be described in a metric framework as GR sourced by a perfect 
fluid with constant energy density $\rho$ and equation of state $p=-\rho$. 
The generation of baryonic matter and the reaction of the condensate allows 
for an effective description in which the fluid becomes anisotropic.
In this effective fluid description, the dark acceleration $a_{\DF}$ is completely due 
to the pressure of the anisotropic fluid $\ppar_{\DF}(r)$.
The (modulus of the) total acceleration experienced by a test particle is given
by~\cite{Cadoni:2017evg}
\begin{equation}
\label{poi} 
\ab + a_{\DF}
\simeq
\frac{\gn\,\mb(r)}{r^2}
+4 \,\pi\, \gn\, r\, \ppar_{\DF}(r)
 \ .
\end{equation} 
Solving Einstein equations sourced by the anisotropic fluid, one finds the 
space-time metric in the form
\be
\label{fluid:metric}
\rmd s^2
=
-f(r) \e{\gamma(r)} \rmd t^2 + \frac{\rmd r^2}{f(r)} +r^2 \rmd \Omega^2
\ ,
\ee
with
\be
f(r)
=
1-\frac{2\,\gn\,m(r)}{r}
\ ,
\ee
and the metric function $\gamma$ determined by the distribution of 
baryonic matter $\mb$,
\be
\label{explicitgammaprime}
\gamma'
=
\frac{2}{r\,f(r)}
\left[\gn\,m_{\rm B}'(r)
+\sqrt{a_0\,\gn\,m_{\rm B}(r)}
\right]
\ .
\ee
It has also been shown that, when we can approximate the baryonic 
mass distribution with a constant profile $m_{\rm B}(r)= m_{\rm B}$ 
(this holds when we consider a galaxy at distances much bigger 
than its bulk), we recover the typical logarithmic behaviour of the 
MOND gravitational potential \cite{Cadoni:2017evg} and Tully-Fisher relation. 
\subsection{Cosmic Balance}
If one puts together the argument based on the graviton number of Section~\ref{ss:cdf}
and the energy balance of Section~\ref{ss:eb}, the ratio between an
{\sl apparent dark matter} mass distribution and baryonic matter can be estimated
and shown to be consistent with the predictions of the $\Lambda$CDM model.
\par
Let us denote with $U_{\DF}$ the energy associated with the dark gravitons. 
This energy can be written in terms of the number $N_{\DF}$ of ``dark force'' 
gravitons inside a sphere of radius $r$ and their Compton energy $\varepsilon =-\mpl \, \lp/r$  
as
\be 
U_{\DF}
=
N_{\DF}\,\varepsilon
=
-N_{\DF} \, \frac{\mpl \, \lp}{r}
\ .
\ee
In the $\Lambda$CDM description, $U_{\DF}$ must be seen as originating from 
the interaction of an {\sl apparent dark matter} mass $M_{\rm DM}$
with the baryonic matter of mass $m_{\rm B}$ and its self-interaction, that is
\be
\label{ppp}
 U_{\DF}
 =
 - \frac{\gn\, m_{\rm DM} \, m_{\rm B}}{r}
 - \frac{\gn \,m_{\rm DM}^2}{r}
 \ .
 \ee
Equating the above two expressions for $U_{\DF}$, we get  
\be
\label{opl}
N_{\DF}
=
\frac{m_{\rm DM}^2}{\mpl^2}
+\frac{m_{\rm DM}\,m_{\rm B}}{\mpl^2}
\ .
\ee
Let us stress that the apparent dark matter mass $m_{\rm DM}$ must 
not be confused with the {\sl effective dark force mass} $m_{\DF}$ of 
Eq.~\eqref{masses}.
In fact, consistency of Eq.~\eqref{opl} with Eq.~\eqref{holoNDF} requires
$m_{\DF}^2={m_{\rm DM}^2}+m_{\rm DM}\,m_B$.
\par 
On using Eq.~\eqref{N0},  $N_{\rm DF}\sim N_{\Lambda} - N_{\rm DE}$,
and recalling that $N_{\Lambda}\sim {L^2}/{\lp^2}$ and $m_{\Lambda} = \mpl\, L / \lp$,
we obtain
\be
m_{\rm DM} \, m _{\rm B} + m_{\rm DM}^2
=
2 \, m_{\Lambda} \, m_{\rm B} - m_{\rm B}^2 
\ ,
\ee
which can be written as
\be
x^2 + x +1
=
\frac{2 \, m_{\Lambda}}{m_{\rm B}}
\ ,
\ee
where we defined the ratio $x = m_{\rm DM} / m_{\rm B}$.
In particular, the latter equation is solved by
\be
\frac{m_{\rm DM}}{m_{\rm B}}
=
\frac{\sqrt{8\,(m_{\Lambda}/m_{\rm B}) - 3}-1}{2}
\ .
\ee
If we now recall that observations yield $m_{\rm B} \simeq 0.05 \, m_{\Lambda}$, we 
finally obtain
\be
\frac{m_{\rm DM}}{m_{\rm B}}
\simeq
5.77
\ ,
\ee
which is in the right ballpark of the $\Lambda$CDM prediction for the 
present relative abundance of dark and baryonic matter.
\section{Conclusions}
\label{sec5}
In this work, we have investigated the emergent laws of gravity by 
modelling our dark energy dominated universe as a critical BEC 
with a large number $N_{\rm G}$ of soft gravitons.
We have shown that the local behaviour of this DEC requires, besides the
usual holographic regime, an extensive regime of gravity in which the graviton
number scales with the volume of space.
Baryonic matter fits naturally in this description as gravitons pulled out from
the DEC at the matter clumping scale give rise to the local (Newtonian)
gravitational forces.
We have then shown that, in this framework, the galaxy rotation curves far away from
the galactic center [{\em i.e.}~the MOND formula~\eqref{MOND}] can likewise
be derived from the reaction of the DEC to the presence of baryonic matter,
without assuming the existence of any sort of dark matter.
We have also evaluated the mass ratio of the apparent dark matter and baryonic
component and found it in agreement with the prediction of the $\Lambda$CDM
model.
\par 
Moreover, our microscopic description can be easily used to produce an emergent theory
of gravity in the form of GR sourced by an anisotropic fluid, the latter being the
macroscopic manifestation of the DEC and of its interaction with baryonic matter.
This can be done along the lines of Refs.~\cite{Faber:2005xc,Cadoni:2017evg},
in which the dark force explaining galactic dynamics furthermore takes the form of a pure
pressure term~\cite{Cadoni:2017evg}.
\par
We would like to conclude by remarking that two important points have not yet
been addressed, but deserve further investigation.
The first one concerns the microscopic origin of the cosmological evolution.
Our model applies solely to the present dark energy dominated universe.
We did not tackle the problem of giving a description of the history of the
universe using a critical BEC of soft gravitons.
Although this is a quite involved problem, there are several indications that it
may indeed be possible.
The results of Refs.~\cite{Dvali:2013eja,Casadio:2015xva,Casadio:2017twg}
about the description of inflation and general cosmological space-times~\cite{Binetruy:2012kx}
represent promising steps along this direction.
Moreover, the results of Ref.~\cite{Carroll:2017kjo} not only assert 
that the dS space-time necessarily appears at late times in any cosmological 
evolution consistent with the generalized second law of thermodynamics, but also 
imply that the presence of an extensive, volume-scaling, term for the graviton 
number is perfectly consistent with this late time cosmological evolution.
Last but not least, the fact that our model predicts the correct present relative
abundance of the various forms of matter gives us a further hint that we are 
going in the right direction. 
\par
The second point concerns the microscopic origin of horizons.
Most of the scenarios for emergent gravity assume in an explicit or implicit way
the presence of event, cosmological or acceleration, horizons
(see, {\em e.g.}~Ref.~\cite{Cadoni:2017qza}). 
Horizons are a key ingredient for explaining the holographic regimes of gravity 
and play, therefore, a crucial role also in our BEC description of black 
holes an the dS space-time.
At the level of the BEC, one may easily generate acoustic horizons~\cite{Barcelo:2000tg}.
However, it is not clear if acoustic horizons in a BEC can be directly linked to
space-time horizons in the emergent gravity scenario.
In fact, acoustic horizons in BEC are mainly of kinematic origin, whereas
in an emergent gravity theory containing black holes and the dS space-time,
their origin should be dynamical.  
\section*{Acknowledgements}
We thank W.~M\"uck for many useful discussions.
A.G.~is grateful to L.~Berezhiani and A.~Giugno for stimulating discussions.
This research was partially supported by INFN, research initiatives FLAG
(R.C.~and A.G.) and QUAGRAP (M.C.~and M.T.).
The work of R.C.~and A.G.~has been carried out in the framework of GNFM
and INdAM and the COST action {\em Cantata\/}.
\bibliographystyle{utphys}
\bibliography{gravitons}
%
\end{document}